\documentclass[aps,pre,amsmath,singlecolumn,showpacs,11pt]{revtex4-2}

\usepackage{amsmath}
\usepackage{amsfonts}
\usepackage{graphicx}
\usepackage{hyperref}
\usepackage{enumerate}
\usepackage{mathtools}
\usepackage{verbatim}
\usepackage{epstopdf}
\usepackage{subfigure}
\usepackage{latexsym}
\usepackage{dcolumn}
\usepackage{comment}
\usepackage{epsf}
\usepackage{float}
\DeclareMathSymbol{\mlq}{\mathord}{operators}{``}
\DeclareMathSymbol{\mrq}{\mathord}{operators}{`'}
\DeclarePairedDelimiter\abs{\lvert}{\rvert}

\usepackage{color}

\usepackage{xcolor} 

\begin{document}
\title{Mean-field theory for double-well systems on degree-heterogeneous networks}
\author{Prosenjit Kundu$^1$} 
\author{ Neil G. MacLaren$^1$} 
\author{Hiroshi Kori$^2$}
\author{Naoki Masuda$^{1,3}$}
\email{naokimas@gmail.com}

\affiliation{$^1$Department of Mathematics, State University of New York at Buffalo, NY 14260-2900, USA }
\affiliation{$^2$Department of Complexity Science and Engineering, The University of Tokyo, Chiba 277-8561, Japan}
\affiliation{$^3$Computational and Data-Enabled Science and Engineering Program, State University of New York at Buffalo, Buffalo, NY 14260-5030, USA}
\affiliation{$^4$Faculty of Science and Engineering, Waseda University, Tokyo 169-8555, Japan}
\date{\today}

\begin{abstract}
Many complex dynamical systems in the real world, including ecological, climate, financial, and power-grid systems,
often show critical transitions, or tipping points, in which the system's dynamics suddenly transit into a qualitatively different state.
In mathematical models, tipping points happen as a control parameter gradually changes and crosses a certain threshold.
Tipping elements in such systems may interact with each other as a network, and understanding the behavior of interacting tipping elements is a challenge because of the high dimensionality originating from the network. Here we develop a degree-based mean-field theory for a prototypical double-well system coupled on a network with the aim of understanding coupled tipping dynamics with a low-dimensional description. The method approximates both the onset of the tipping point and the position of equilibria with a reasonable accuracy. Based on the developed theory and numerical simulations, we also provide evidence for multistage tipping point transitions in networks of double-well systems.
\end{abstract}

 \maketitle
\section{Introduction\label{sec:introduction}}

Many empirical complex systems show nonlinear dynamics. Nonlinear behavior sometimes leads to a situation in which the system suddenly shifts from one stable state to a qualitatively, or drastically, different stable state. Reversing such shifts may be difficult due to hysteresis or other reasons. This behavior is widely known as a critical transition or tipping point \cite{Scheffer_book2009%
%
%
}. Tipping points are observed in a variety of complex dynamical systems including ecological, biological, social, and economical systems. Examples include nutrient-driven shifts in shallow lakes between clear and turbid water \cite{%
%
%
Carpenter_Science2011}, abrupt climate changes \cite{Lenton_PNAS2008},  epileptic seizures \cite{Litt_Neuron2001},
and extensive crashes in financial markets \cite{May_Nature2008}.

Many tipping point phenomena in nature can be modeled by complex systems that are a collection of tipping elements interacting through a network.
Examples include extinction of species in ecosystems in which different species are connected by foodweb, mutualistic, and other ecological networks   \cite{Allesina_Nature2012,
%
%
Grilli_NatComm2017}, blackouts in power grids \cite{Carreras_Chaos2002,Dobson_Chaos2007
%
%
}, psychiatric disorders \cite{Hayes_Clin_Phys_Rev2015,Cramer_PlosOne2016%
%
%
} and the Greenland ice sheet \cite{Kriegler_PNAS2009,Klose_RSOS2020}.
Obviously, the behavior of the tipping elements is not independent of each other when they are connected as a network \cite{Scheffer_Nature2009,Scheffer_Science2012, Boettiger_Nature2013}.
%
%
 For instance, suppose that a tipping element, denoted by $v$, passes the tipping point, i.e., the value of a control parameter of the system at which a tipping point occurs. Then, it is more likely that a second tipping element directly connected with $v$ will tip as well, resulting in a domino effect \cite{Kinzig_EcoSoc2006, Rocha_Science2018}.

Accurately describing the tipping points in dynamics on networks is desirable because we may want to anticipate and prevent them as well as to know the magnitude of a tipping cascade in the network. However, high dimensionality of dynamical systems on a network and the complex structure of the network at hand generally make this task difficult.
One strategy towards this goal is to reduce the dimension of the original system without loss of too much information about the original dynamics. For example, Gao, Barzel, and Barab\'{a}si proposed a theory to reduce a class of dynamical system on networks, including networks of dynamical tipping elements, into one-dimensional dynamics~\cite{Gao_Nature2016}. This method, which we refer to as GBB reduction, is applicable to directed and weighted networks and uses each node's weighted in-degree (i.e., the number of incoming edges) and out-degree (i.e., the number of outgoing edges). The GBB reduction has been generalized to the methods  which  use the eigenvalues and eigenvectors of the adjacency matrices or related matrices of the network \cite{Laurence_PRX2019, Thibeault_PRR2020, Masuda_PRR2022}. A related but different two-dimensional reduction locates the tipping point with a high accuracy for ecological dynamics occurring on bipartite networks representing mutualistic interactions~\cite{Jiang_PNAS2018}.

The GBB reduction is a degree-based mean-field (DBMF) theory, also called heterogeneous mean-field theory, in the sense that the theory only uses the information about the degree of each node. In general, a DBMF theory for network dynamics assumes that nodes with the same degree are statistically the same and obey the same dynamical rule. It is expected to be accurate for large uncorrelated random networks with general degree distributions. In fact, the most common family of DBMF theory is different from the GBB reduction in that the former aims at deriving a dynamical equation that any node with a given degree obeys, one equation per degree value. In contrast, the goal of the GBB reduction and its extensions is to derive a one-dimensional or low-dimensional reduction of the entire dynamics on the network. To avoid confusion, we save the term DBMF theory only to refer to the conventional type of the DBMF theory and do not refer to the GBB reduction as a DBMF theory below.
The DBMF theory (in the conventional sense) has been successful in describing percolation \cite{Cohen_PRL2000,Callaway_PRL2000,Cohen_PRL2001, Newman_PRE2001, Newman_PRE2002%
%
%
}, dynamic epidemic processes \cite{Pastor_PRL2001,Pastor_PRE2001,Pastor_PRE2002a,Boguna_PRE2002,Moreno_EPJB2002, Masuda_JTB2006,Barrat_dynamical_book,%
%
%
Pastor_RMP2015}, synchronization \cite{Ichinomiya_PRE2004,Coutinho_PRE2013, Kundu_PRE2017, Kundu_Chaos2019}, random walks \cite{Baronchelli_PRE2010, Lau_EPL2010}, voter models \cite{Antal_PRE2006, Sood_PRE2008}, and rock-scissors-paper games \cite{Masuda_PRE2006}, to name a few, on degree-heterogeneous uncorrelated networks.
A similar DBMF theory for networks of tipping elements, if accurate, will be a useful tool for anticipating tipping points as a system parameter gradually changes and for understanding the impact of the degree distribution on the tipping behavior.
Thus motivated, in the present study, we develop a DBMF theory for prototypical dynamical systems showing tipping behavior, i.e., the double-well system, connected as networks.

\section{Model}

We consider networks in which each node is a tipping element that obeys deterministic dynamics of a double-well system.
When there is no interaction between nodes, the state of each node evolves according to
\begin{equation}
\dot{x} = - (x - r_1) (x - r_2) (x - r_3)  + u,
\label{eq:double_well1}
\end{equation}
where $r_1$, $r_2$, and $r_3$ are constants (with $r_1 < r_2 < r_3$) and $u$ represents an external input applied to the node. 
This system has a unique stable equilibrium, $x^*$, with $x^* < r_1$ when $u < u_{\rm c, \ell}$, where $u_{\rm c, \ell}$ is schematically shown in
Fig.~\ref{cubic_poly_ss}. 
We call this equilibrium the lower equilibrium and denote it by $x^{\rm \ell}$.
The system has a unique stable equilibrium satisfying $x^* > r_3$, when $u > u_{\rm c, u}$, where $u_{\rm c, u}$ is shown in Fig.~\ref{cubic_poly_ss}. 
We call this equilibrium the upper equilibrium and denote it by $x^{\rm u}$.
Both $x^{\rm \ell}$ and $x^{\rm u}$ are stable when $u\in (u_{\rm c, \ell}, u_{\rm c,u})$. 

If we initialize the system at the lower equilibrium and gradually increase $u$, then the system undergoes a saddle-node bifurcation at $u=u_{\rm c, u}$ such that $x^*$ jumps from the lower equilibrium to the upper equilibrium.  If we initialize the system with the upper equilibrium and gradually decrease $u$, then the system jumps to the lower equilibrium at $u=u_{\rm c, \ell}$, implying a hysteresis. Similar models have been used for 
representing alternative stable states and hysteresis in ecosystems~\cite{Beisner_FEE2003,Wunderling_Chaos2020,Brummitt_JRSI2015,Wunderling_ESD2021}, thermohaline circulation  \cite{Wright_JPO1991}, and ice sheets \cite{Levermann_TCryo2016}.

We consider tipping elements that obey the coupled double-well system dynamics given by 
\begin{equation}
\dot{x}_i = - (x_i - r_1) (x_i - r_2) (x_i - r_3) + D \sum_{j=1}^N A_{ij} x_j +u\quad (i\in \{1, \ldots, N\}),
\label{eq:double_well}
\end{equation}
where $N$ is the number of nodes, $D$ is the coupling strength, and $A = (A_{ij})$ is the $N\times N$ adjacency matrix of the network.
Equation~\eqref{eq:double_well} has been used for modeling ecological systems such as lake chains in which a node represents a lake~\cite{Klose_RSOS2020}, global interactions among climate tipping elements such as the Greenland ice sheet, the Atlantic Meridional Overturning Circulation (AMOC), and the Amazon rainforest \cite{Wunderling_ESD2021}, and Amazon as a network of tipping elements in which each node represents forests within a specific area \cite{Wunderling_Chaos2020} (also see Refs.~\cite{Brummitt_JRSI2015, Kronke_PRE2020, Wunderling_NJP2020} for analyses of the same model). For simplicity, we consider undirected and unweighted networks such that $A_{ij} = A_{ji} \in \{ 0, 1 \}$ for any $i, j \in \{1, \ldots, N\}$. We exclude self-loops, i.e., we set $A_{ii} = 0$ $\forall i\in \{1, \ldots, N\}$. We denote by $k_i =\sum_{j=1}^N A_{ij} = \sum_{j=1}^N A_{ji}$ the degree of the $i$th node and by $p(k)$ the degree distribution. This model allows us to investigate  tipping cascades on networks.

\section{Degree-based mean-field theory\label{sec:mean-field}}

To describe the dynamics given by Eq.~\eqref{eq:double_well}, we develop a DBMF theory, which assumes that the nodes with degree $k$ are statistically equivalent to each other and that the states of different nodes are statistically independent of each other except their dependence on $k$. This assumption corresponds to a configuration model of networks, i.e., a random network with a given degree sequence or distribution. In this theory, we replace the elements of the adjacency matrix, $A_{ij}$, by its ensemble average, denoted by $\langle A_{ij} \rangle$, which represents the probability that the $i$th and $j$th nodes are adjacent to each other, assuming $\langle A_{ij} \rangle \le 1$.
For uncorrelated networks, we obtain
\begin{equation}
\langle A_{ij}\rangle= \frac{k_ik_j}{N\langle k \rangle},
\label{annealed2}
\end{equation}
where $\langle k \rangle$ is the average degree.
Equation~\eqref{annealed2} implies that 
the probability that the $j$th node is a neighbor of an arbitrary node is equal to $k_j/\left(N\langle k\rangle \right)$.
The DBMF theory based on Eq.~\eqref{annealed2} was first pioneered for percolation
and the susceptible-infectious-susceptible model (see Section~\ref{sec:introduction} for references).

By substituting Eq.~\eqref{annealed2} in Eq.~\eqref{eq:double_well}, we obtain
\begin{equation}
\dot{x}_i = - (x_i - r_1) (x_i - r_2) (x_i - r_3) + D k_i \Theta +u,
\label{eq:double_well_hmf}
\end{equation}
where
\begin{equation}
\Theta = \frac{1}{N\langle k\rangle}
\sum_{j=1}^N k_j x_j
\label{self_con1}
\end{equation}
is the degree-weighted average of $x_j$.

\begin{figure}
\centering
\includegraphics[height=!,width=0.6\textwidth]{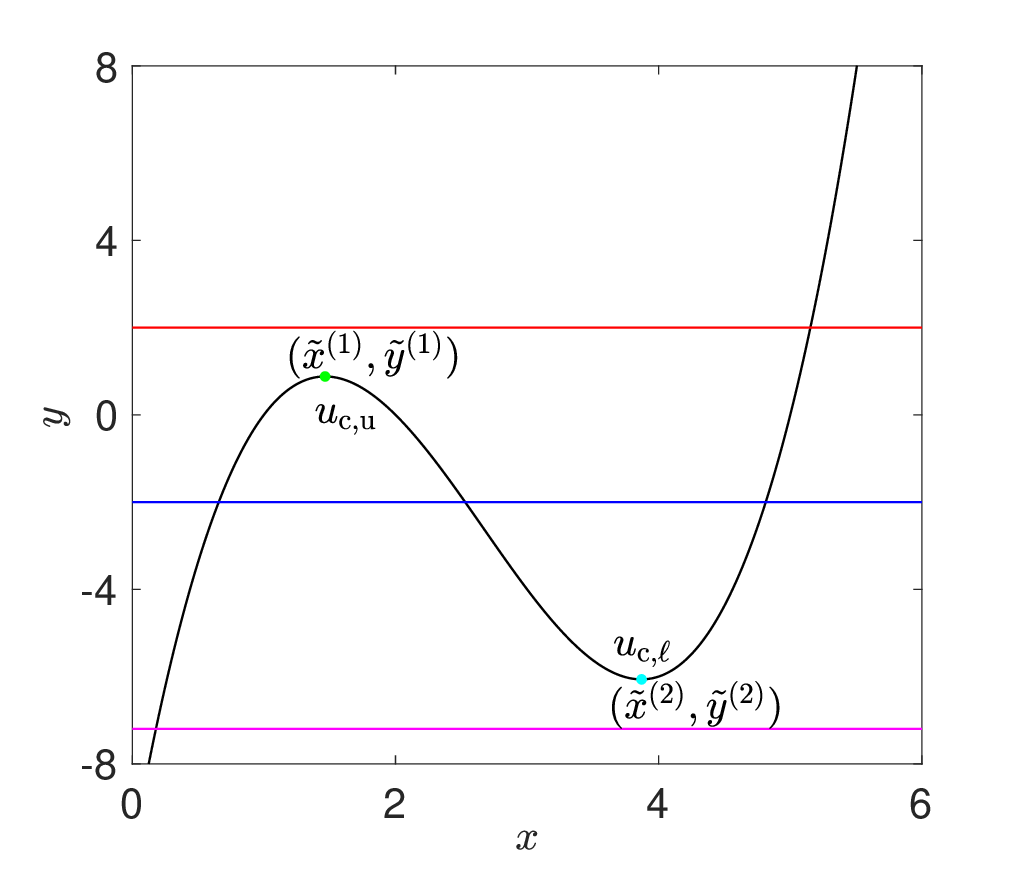}
\caption{Graphical representation of the equilibria of the single and the coupled double-well systems. Under the DBMF theory, the intersection of $y = (x_i-r_1)(x_i-r_2)(x_i-r_3)$ and $y = Dk_i\Theta + u$ gives $x_i$ at equilibrium.}
\label{cubic_poly_ss}
\end{figure}

For a given $\Theta$ value, the equilibria of
Eq.~\eqref{eq:double_well_hmf} are given by the intersection of the cubic polynomial
$y = (x_i - r_1) (x_i - r_2) (x_i - r_3)$ and the horizontal line $y = D k_i \Theta + u$ in the $x$-$y$ plane (see Fig.~\ref{cubic_poly_ss}). Denote by $(\tilde{x}^{(1)}, \tilde{y}^{(1)})$ the unique local maximum of
$y = (x - r_1) (x - r_2) (x - r_3)$ and by $(\tilde{x}^{(2)}, \tilde{y}^{(2)})$ the unique local  minimum.

For the nodes with the smallest degrees satisfying
$D k_i \Theta + u < \tilde{y}^{(2)}$, 
the curve $y = (x_i - r_1) (x_i - r_2) (x_i - r_3)$ and the line $y = D k_i \Theta + u$ have just one intersection at $x_i < \tilde{x}^{(1)}$ such that
there is a unique stable equilibrium, which we call the lower state (see the magenta line 
in Fig.~\ref{cubic_poly_ss}).
For the nodes with the largest degrees satisfying
$D k_i \Theta + u > \tilde{y}^{(1)}$, there is a unique stable equilibrium satisfying $x_i \ge x^{(2)}$, which we call the upper state (see the red line in Fig.~\ref{cubic_poly_ss}).
Finally, for the nodes whose degree satisfies $\tilde{y}^{(2)} < D k_i \Theta + u < \tilde{y}^{(1)}$,
Eq.~\eqref{eq:double_well_hmf} allows both lower and upper states (see the blue line in Fig.~\ref{cubic_poly_ss}).

We assume that all nodes are initially in their respective lower states and that we gradually increase $u$.
A larger value of $D k_i \Theta + u$ induces the upper state for the $i$th node under the DBMF theory.
Therefore, at a given value of $u$, the $Nq$ nodes with the smallest degree, i.e., the nodes whose $k_i$ is smaller than a threshold value $\tilde{k}$, are in the lower state, where $q$ is the fraction of nodes in the lower state. The other $N(1-q)$ nodes, which have $k_i \ge \tilde{k}$, are in the upper state.
The value of $k_i$ that satisfies $D k_i \Theta + u = \tilde{y}^{(1)}$  is equal to $\tilde{k}$. In other words, we obtain
\begin{equation}
\tilde{k} = \frac{\tilde{y}^{(1)} - u}{D\Theta^*},
\label{eq:tilde_k}
\end{equation}
where $\Theta^*$ is the $\Theta$ value in the equilibrium.

Similar to the analysis of epidemic process models~\cite{Pastor_PRL2001,Pastor_PRE2001,Pastor_PRE2002a,Boguna_PRE2002,Moreno_EPJB2002,Barrat_dynamical_book,Pastor_RMP2015, Masuda_JTB2006}, we need to self-consistently determine $\Theta^*$.  
Using Eq.~\eqref{self_con1}, we obtain
\begin{equation}
\Theta^* = \frac{1}{N\langle k\rangle}
\left[
\sum_{k < \tilde{k}} k p(k) x^{\rm \ell}(k) +
\sum_{k \ge \tilde{k}} k p(k) x^{\rm u}(k)
\right].
\label{eq:self_consistenta_Theta^*}
\end{equation}
where $x^{\rm \ell}(k)$ and $x^{\rm u}(k)$ are the lower and upper states, respectively, which are functions of $k$ and $\Theta^*$.

We obtain $\Theta^*$ and $\tilde{k}$ from Eqs.~\eqref{eq:tilde_k} and \eqref{eq:self_consistenta_Theta^*} as follows.
First, the right-hand side of Eq.~\eqref{eq:self_consistenta_Theta^*}, which we write $\Theta^*_2(\tilde{k})$, is a continuous and monotonically decreasing function of $\tilde{k}$, given the $\Theta^*$ value with which to calculate the $x^{\rm \ell}(k)$ or $x^{\rm u}(k)$ value for each $k$. We obtain
$\lim_{\tilde{k}\to 0} \Theta^*_2(\tilde{k}) = \Theta^*_2(k_{\min}) = \langle k x^{\rm u}(k) \rangle / \langle k\rangle$ and $\lim_{\tilde{k}\to\infty} \Theta^*_2(\tilde{k}) = \lim_{\tilde{k} \downarrow k_{\max}} \Theta^*_2(\tilde{k}) = \langle k x^{\rm \ell}(k) \rangle / \langle k \rangle$, where $\langle \cdot \rangle$ is the average over the $N$ nodes.
Second, we rewrite Eq.~\eqref{eq:tilde_k} as $\Theta^*_1(\tilde{k}) \equiv (\tilde{y}^{(1)} - u)/(D\tilde{k})$. It holds true that
$\Theta^*_1(\tilde{k})$ is also a continuous and monotonically decreasing function of $\tilde{k}$, $\lim_{\tilde{k} \downarrow 0} \Theta^*_1(\tilde{k}) = \infty$, and
$\lim_{\tilde{k}\to\infty} \Theta^*_1(\tilde{k}) = 0$. Because $\lim_{\tilde{k}\downarrow 0} \left[ \Theta^*_1(\tilde{k}) - \Theta^*_2(\tilde{k}) \right] > 0$ and
$\lim_{\tilde{k}\to\infty} \left[ \Theta^*_1(\tilde{k}) - \Theta^*_2(\tilde{k}) \right] < 0$, there is at least one positive value of $\tilde{k}$ that solves
$\Theta^*_1(\tilde{k}) - \Theta^*_2(\tilde{k}) = 0$, which one can obtain by, for example, the bisection method. Although there may be multiple roots
of $\Theta^*_1(\tilde{k}) - \Theta^*_2(\tilde{k}) = 0$ depending on the degree distribution, with a practical degree distribution,
 $\Theta^*_1(\tilde{k}) - \Theta^*_2(\tilde{k}) = 0$ usually has only one root. We regard that all the nodes are in the upper or lower states if the obtained solution satisfies $\tilde{k} \le k_{\min}$ or $\tilde{k} > k_{\max}$, respectively.

\section{Numerical results\label{sec:numerical}}

In this section, we mainly investigate the accuracy of the DBMF theory developed in Section~\ref{sec:mean-field} against direct numerical simulations in locating the tipping point and approximating $\{ x_1, \ldots, x_N \}$ at equilibrium. 
\subsection{Numerical methods and networks}

We set $r_1 = 1$, $r_2 = 2$, and $r_3 = 5$. We either fix $D$ and vary $u$ (Section~ \ref{sub:vary-u}) or fix $u$ and vary $D$ (Section~\ref{sub:vary-D}).
For the given $u$ and $D$ values,
we set the initial conditions to $x_i=0.01$ (with $i=1,\dots,N$) and run simulations for 30000 time units to ensure that the equilibrium has been reached.
At a relatively small value of $u$ or $D$, all the nodes are in their lower state at equilibrium. Then, we gradually increase $u$ or $D$.

We use the following networks. First, we use networks generated by the Barab{\'a}si-Albert (BA) model \cite{Barabasi_Science1999} that produces scale-free networks with $p(k) \propto k^{-3}$, where $\propto$ represents ``proportional to''. We refer to these networks as BA networks. To generate a BA network, we start with a network with two nodes connected by an edge and connect each of the two edges emanating from each new node to an existing node according to the linear preferential attachment rule. We obtain $\langle k\rangle \approx 4$, where $\approx$ represents ``approximately equal to''.

Second, we use the Holme-Kim network model \cite{Holme_PRE2002}, which is a modification of the BA model for high clustering (i.e., a large number of triangles). We construct networks with average degree $\langle k \rangle \approx 10$ by setting the number of edges that each new node brings in to five. We set the probability of constructing a triangle for the each new edge to $0.5$.
Third, we use the largest connected component (LCC) of a coauthorship network of researchers in network science, which has $N = 379$ nodes and 914 undirected edges \cite{Newman_PRE2006}. Each node in this network represents a researcher publishing a paper in network science up to year $2006$. An edge exists between two nodes if the two researchers coauthored at least one paper.
Finally, we use the LCC of the hamsterster social network, which has $N = 1788$ nodes and 12476 undirected edges \cite{Kunegis_PICWWW2013}. A node in this network represents a user of hamsterster.com. Two users are adjacent if they have a friendship relation on the website.

We compare the performance of the DBMF theory with that of the GBB  reduction.
The GBB reduction for the coupled double-well system (see Eq.~\eqref{eq:double_well}) is given by 
\begin{equation}
\dot{x} = - (x - r_1) (x - r_2) (x - r_3) + D \beta x +u,
\label{GBB1D}
\end{equation}
where $\beta=\sum_{i=1}^N k_i^2 \big/ \sum_{i=1}^N k_i$ \cite{Gao_Nature2016}. 
When the GBB reduction is accurate, the dynamics and equilibria of $x$ obtained from Eq.~\eqref{GBB1D} closely approximate those of the observable 
$x_{\rm eff}$, called the effective state, given by
\begin{equation}
x_{\rm eff}=\frac{\sum_{i=1}^N k_ix_i}{\sum_{i=1}^N k_i},
\label{x_eff1}
\end{equation}
where we obtain $\{x_1, \ldots, x_N\}$ from direct numerical simulations of the original $N$-dimensional dynamical system, Eq.~\eqref{eq:double_well}. 
To compare the accuracy of the DBMF theory with the GBB, we measure the effective state calculated from
the DBMF theory given by
\begin{eqnarray}
x_{\rm DBMF}&=&\frac{\sum_{k = k_{\min}}^{k_{\max}} kx(k)}{\sum_{k = k_{\min}}^{k_{\max}} k},
\label{x_MF}
\end{eqnarray}
where $x(k)$ is either $x^{\ell}(k)$ or $x^{\rm u}(k)$.


\subsection{Tipping under gradual increases in $u$\label{sub:vary-u}}

\begin{figure}
\centering
\includegraphics[height=!,width=\textwidth]{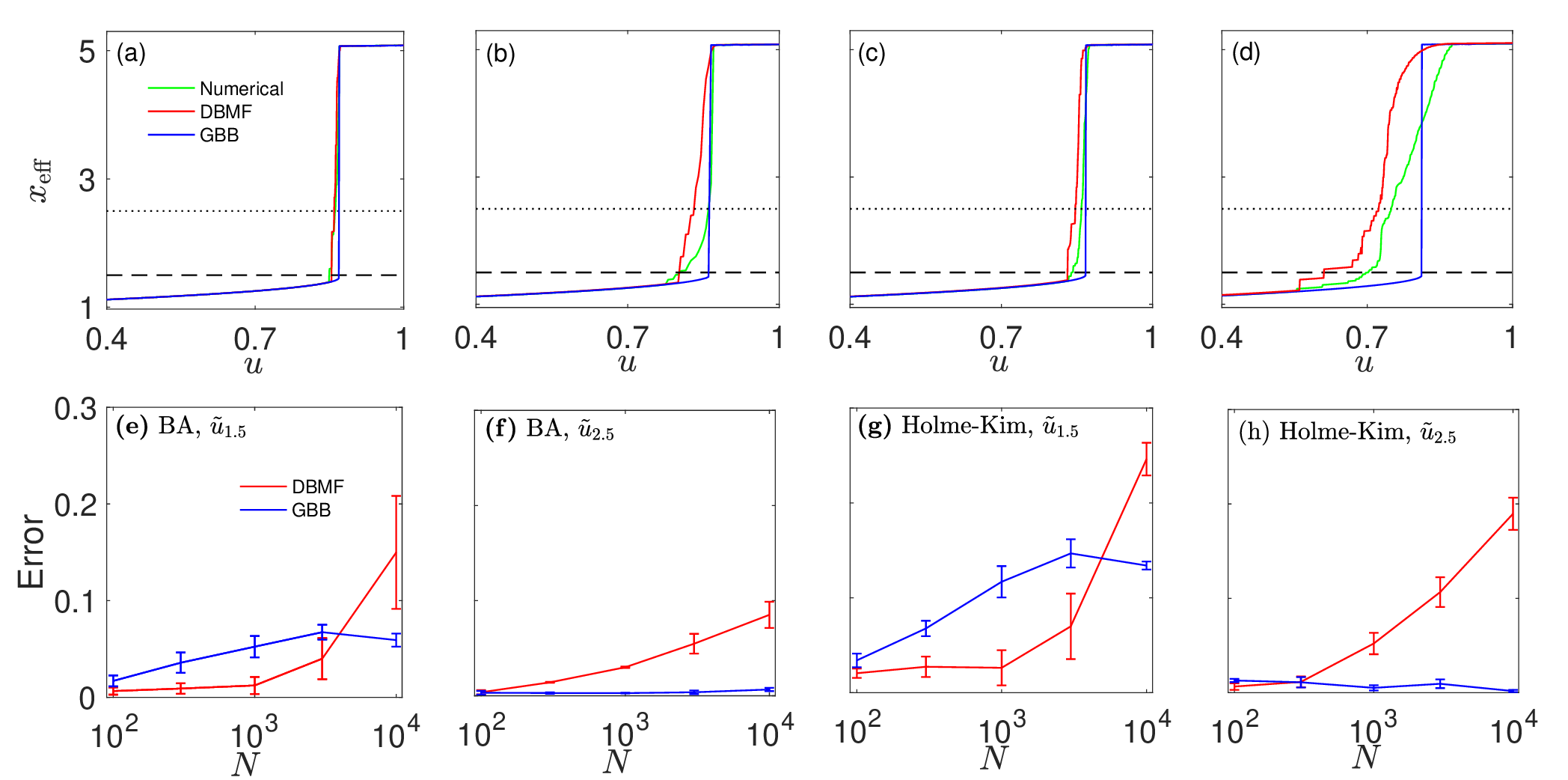}
\caption{Performance of the DBMF theory and GBB reduction on approximating the effective state, $x_{\rm eff}$, in different networks when we set $D=0.001$ and vary $u$. (a) BA network with $N=10^2$ nodes. (b) BA network with $N=10^3$ nodes. (c) Coauthorship network. (d) Hamsterster network. (e) Approximation error for the DBMF theory and GBB reduction in locating the tipping point, $\tilde{u}_{1.5}$, for BA networks with different numbers of nodes.
 (f) Approximation error in locating $\tilde{u}_{2.5}$ for BA networks. (g) Approximation error in locating $\tilde{u}_{1.5}$ for Holme-Kim networks. (h) Approximation error in locating $\tilde{u}_{2.5}$ for Holme-Kim networks.
In (a)--(d), the dashed and dotted lines represent $x_{\rm eff} = 1.5$ and $x_{\rm eff} = 2.5$, respectively.
In (e)--(h), the error bars represent the average $\pm$ standard deviation calculated on the basis of 100 simulations, each of which uses a different network instance with the given number of nodes. } 
\label{Bif_point_prediction}
\end{figure}

We set $D = 0.001$ and vary $u$ in this section.
We show the dependence of $x_{\rm eff}$ on $u$ for a BA network with $N = 10^2$ nodes, a BA network with $N=10^3$ nodes, the coauthorship network, and the hamsterster network in 
Figs.~\ref{Bif_point_prediction}(a), \ref{Bif_point_prediction}(b), \ref{Bif_point_prediction}(c), and \ref{Bif_point_prediction}(d), respectively. 
We find that, in numerical simulations, the tipping point does not occur at the same value of $u$ for the different nodes, resulting in gradual transitions from the lower state to the upper state as one increases $u$; we will examine this phenomenon in Section~{\ref{sub:multistage}} . Therefore, we need to define which tipping points we want to anticipate by theory. Because we are usually interested in critical points at which a macroscopic number of nodes (i.e., $O(N)$) starts to experience a major transition, we operationally define the tipping point as the value of $u$ at which $x_{\rm eff}$ exceeds $1.5$, denoted by $\tilde{u}_{1.5}$ (see the dashed lines in Fig.~\ref{Bif_point_prediction}), 
for the first time as we gradually increase $u$.
Roughly speaking, 5--10\%  of the nodes have switched from the lower state to the upper state at $u = \tilde{u}_{1.5}$.
 Figures~\ref{Bif_point_prediction}(a)--(d) suggest that the DBMF theory better approximates $\tilde{u}_{1.5}$  than the GBB reduction does. We quantitatively confirm this claim in Table~\ref{table1}, which compares the error in the tipping point defined by $\abs{u_{\rm GBB} - \tilde{u}_{1.5}}$  for GBB reduction and $\abs{u_{\rm DBMF} - \tilde{u}_{1.5}}$ for our DBMF theory, where $u_{\rm GBB}$ and $u_{\rm DBMF}$ are the tipping point estimated by the GBB reduction and the DBMF theory, respectively. 
We note that we do not add a generalization of the GBB reduction \cite{Laurence_PRX2019,Thibeault_PRR2020} to the present comparison because the DBMF theory and the GBB reduction only use the degree of each node, whereas the generalizations of the GBB reduction use the full adjacency matrix of the network \cite{Masuda_PRR2022}.

The accuracy of the different approximations depends on the definition of the tipping point. To show this, consider a tipping point defined as the value of $u$ at which $x_{\rm eff}$ exceeds $2.5$ for the first time as one increases $u$, denoted by $\tilde{u}_{2.5}$ (see the dotted lines in Fig.~\ref{Bif_point_prediction}). With this definition, the DBMF theory is better at locating the tipping point than the GBB reduction for two networks and vice versa for the other two networks (see Table~\ref{table1}).

We also confirmed that, for BA networks of different sizes, the error in locating the tipping point is systematically smaller for the DBMF theory than the GBB reduction when we use $\tilde{u}_{1.5}$ as the tipping point (see Fig.~\ref{Bif_point_prediction}(e)), whereas the opposite is the case
with $\tilde{u}_{2.5}$ (see Fig.~\ref{Bif_point_prediction}(f)). 
With both $\tilde{u}_{1.5}$ and $\tilde{u}_{2.5}$, the error increases as $N$ increases.

We show the approximation error for the GBB reduction and the DBMF theory for Holme-Kim networks
in Figs.~\ref{Bif_point_prediction}(g) and \ref{Bif_point_prediction}(h) for $\tilde{u}_{1.5}$ and $\tilde{u}_{2.5}$, respectively. 
The results are similar to the case of BA networks (see Figs.~\ref{Bif_point_prediction}(e) and \ref{Bif_point_prediction}(f)) despite the presence of high clustering in the Holme-Kim networks.
Specifically, the DBMF theory is better approximating the tipping point than the GBB reduction for the tipping point $\tilde{u}_{1.5}$ (see Fig.~\ref{Bif_point_prediction}(g)), whereas the opposite holds true except for small networks in the case of $\tilde{u}_{2.5}$ (see Fig.~\ref{Bif_point_prediction}(h)).

\begin{table}[h!]
\centering
    \caption{Approximation error for the DBMF theory and GBB reduction. We set $D=0.001$ and gradually increase $u$ to determine tipping points
    $\tilde{u}_{1.5}$ and $\tilde{u}_{2.5}$. We set $u=0$ and gradually increase $D$ to determine tipping points $\tilde{D}_{1.5}$ and $\tilde{D}_{2.5}$.
    }
    \begin{tabular}{|m{2.5cm}| m{1.2cm}| m{1.2cm}| m{1.2cm}|m{1.2cm}|m{1.2cm}|m{1.2cm}|m{1.2cm}|m{1.2cm}|} 
    \hline
   ~& \multicolumn{2}{|c|}{$ \tilde{u}_{1.5}$} & \multicolumn{2}{|c|}{$\tilde{u}_{2.5}$}  & \multicolumn{2}{|c|}{$\tilde{D}_{1.5}$} & \multicolumn{2}{|c|}{$\tilde{D}_{2.5}$} \\
     \hline	
     Network  &  DBMF& GBB& DBMF& GBB &  DBMF& GBB& DBMF& GBB\\
      \hline
      BA, $N = 10^2$ & 0.004 &  0.014 & 0.002 &  0.007& 0.007 &  0.048 & 0.001 &  0.031\\
       \hline
       BA, $N = 10^3$ & 0.002 &  0.053&  0.026 &  0.002 & 0.012 &  0.034&  0.023 &  0.007\\
       \hline
       Coauthorship  & 0.008 &  0.027 &  0.012 &  0.008& 0.009 &  0.043 &  0.012 &  0.029\\
        \hline
       Hamsterter  & 0.092 &  0.109  & 0.025 &  0.063& 0.001 &  0.007  & 0.001 &  0.005\\
        \hline
    \end{tabular}
     \label{table1}
\end{table}

\begin{figure}
\centering
\includegraphics[height=!,width=0.9\textwidth]{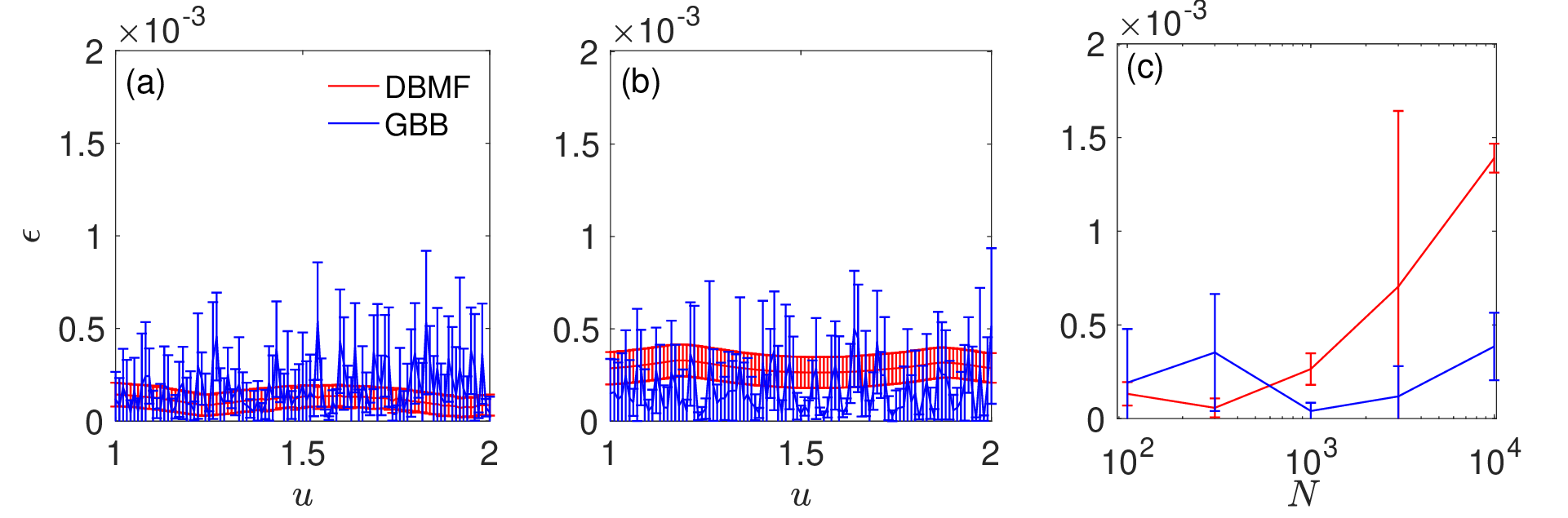}
\caption{Relative error in approximating the effective state by the DBMF theory and GBB reduction. We use BA networks with $N$ nodes.
(a) $N=100$. (b) $N=1000$. (c) Dependence on $N$ at $u=1.5$. 
In (a) and (b), we omitted the part of the error bar below $\epsilon = 0$.}
\label{Curve_predction}
\end{figure}

We next compared the performance of the DBMF theory and the GBB reduction in approximating the 
the effective state, $x_{\rm eff}$, off the tipping point, which has been a main goal of  recent papers on dimension reduction techniques for dynamical systems on networks \cite{Gao_Nature2016, Tu_PRE2017,Laurence_PRX2019,Thibeault_PRR2020, Tu_Iscience2021, Kundu_PRE2022}. 
To quantify the accuracy at approximating $x_{\rm eff}$, we measure the relative error between the results obtained from direct numerical simulations and the theoretical estimate, which we denote by $\epsilon$. For the DBMF theory,
we obtain $\epsilon = \abs{(x_{\rm eff}-x_{\rm DBMF})/x_{\rm eff}}$, where $x_{\rm eff}$ refers to the value obtained from direct numerical simulations.
For the GBB reduction, we set $\epsilon = \abs{(x_{\rm eff}-x{(\beta)})/x_{\rm eff}}$, where we obtain $x(\beta)$ as the equilibrium of the one-dimensional dynamics given by Eq.~\eqref{GBB1D}.
At each $u$ value, we calculate the average and standard deviation of the relative error on the basis of 100 BA networks.
We show the relative error for BA networks with $N=10^2$ and $N=10^3$ nodes in Figs.~\ref{Curve_predction}(a) and \ref{Curve_predction}(b), respectively, for a range of $u$ above the tipping point. 
The relative error for the DBMF theory is similar to that for the GBB reduction when $N=10^2$. The GBB reduction yields smaller errors when $N=10^3$. However, the relative error for the DBMF theory when $N=10^3$ is also fairly small (i.e., $< 0.5 \times 10^{-3}$). We show in Fig.~\ref{Curve_predction}(c) the mean and standard deviation of the relative error at $u=1.5$ as a function of $N$. Although the error for the DBMF theory increases as $N$ increases, it remains reasonably small up to $N=10^4$.

\subsection{Tipping under gradual increases in $D$\label{sub:vary-D}}

\begin{figure}
\centering
\includegraphics[height=!,width=\textwidth]{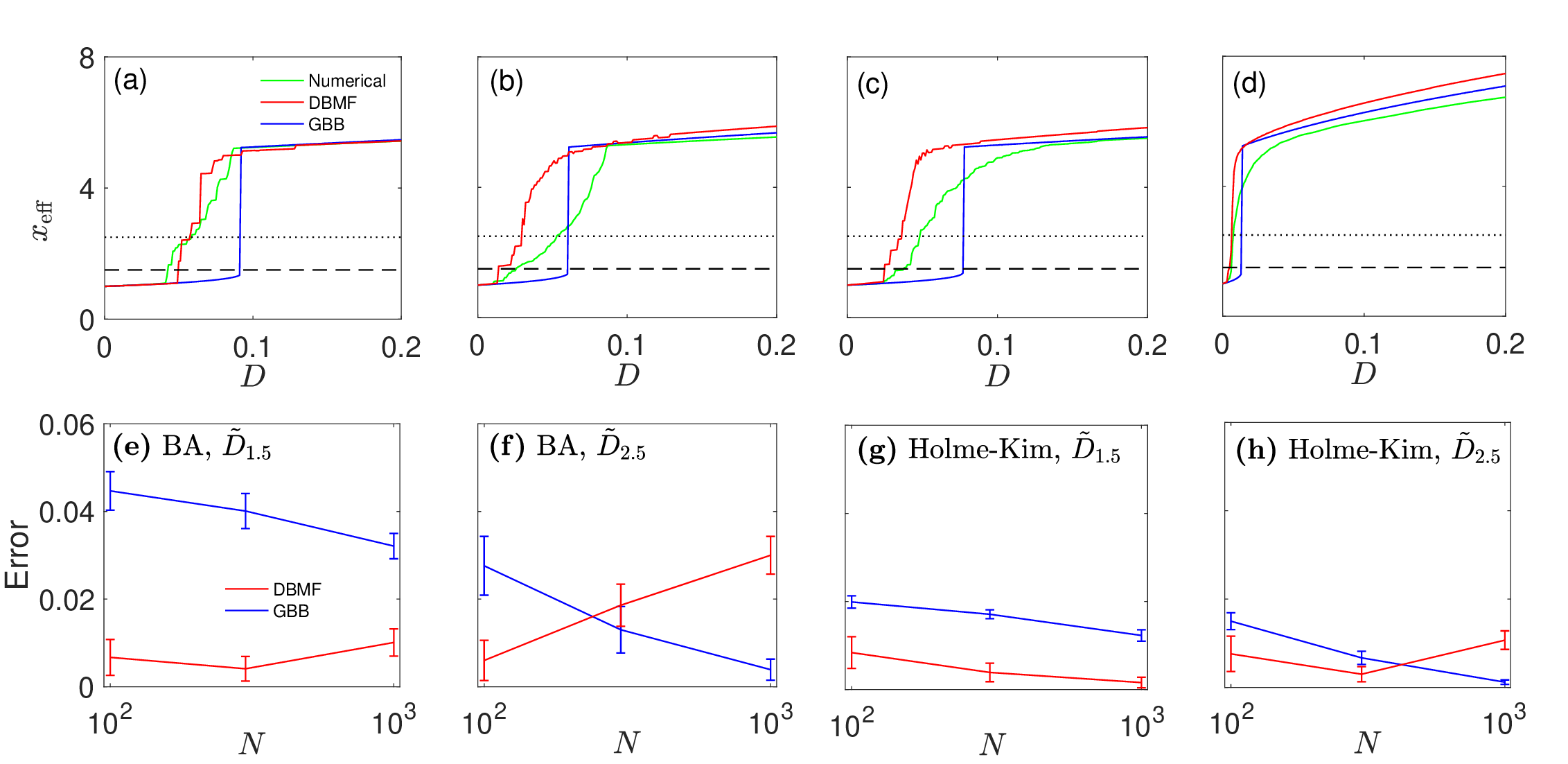}
\caption{Performance of the DBMF theory and GBB reduction on approximating the effective state, $x_{\rm eff}$, in different networks when we set $u = 0$ and vary $D$. (a) BA network with $N=10^2$ nodes. (b) BA network with $N=10^3$ nodes. (c) Coauthorship network. (d) Hamsterster network. (e) Approximation error for the DBMF theory and GBB reduction in locating the tipping point, $\tilde{D}_{1.5}$, for BA networks.
(f) Approximation error in locating $\tilde{D}_{2.5}$ for BA networks. (g) Approximation error in locating $\tilde{D}_{1.5}$ for Holme-Kim networks. (h) Approximation error in locating $\tilde{D}_{2.5}$ for Holme-Kim networks.
In (a)--(d), the dashed and dotted lines represent $x_{\rm eff} = 1.5$ and $x_{\rm eff} = 2.5$, respectively.
See the caption of Fig.~\ref{Bif_point_prediction} for the definition of the error bar. }
\label{Bif_point_varyD}
\end{figure}

Next, we examine the accuracy of the DBMF theory and GBB reduction when we set $u=0$ and vary $D$. In Fig.~\ref{Bif_point_varyD}(a), (b), (c), and (d), we show the relationship between $x_{\rm eff}$ and $D$ for a BA network with $N=10^2$ nodes, a BA network with $N=10^3$ nodes, the coauthorship network, and the hamsterster network, respectively. For approximating the location of the tipping point, we find that our DBMF theory is more accurate than the GBB reduction when we define the tipping point by $\tilde{D}_{1.5}$,
 i.e, the value of $D$ at which $x_{\rm eff}$ exceeds $1.5$ for the first time when we increase the value of $D$ (see the dashed lines in Fig.~\ref{Bif_point_varyD}). If we define the tipping point by $\tilde{D}_{2.5}$ (see the dotted lines in Fig.~\ref{Bif_point_varyD}), the DBMF theory is more accurate on the BA network with $N=10^2$ nodes (Fig.~\ref{Bif_point_varyD}(a)) and the hamsterster network (Fig.~\ref{Bif_point_varyD}(d)) and vice versa
on the BA network with $N=10^3$ nodes (Fig.~\ref{Bif_point_varyD}(b)) and the coauthourship network (Fig.~\ref{Bif_point_varyD}(c)); also see
Table~\ref{table1}.

With BA networks with different numbers of nodes, the DBMF theory locates $\tilde{D}_{1.5}$ more accurately than the GBB reduction (see Fig.~\ref{Bif_point_varyD}(e)). The opposite is the case when we define the tipping point by  $\tilde{D}_{2.5}$  (see Fig.\ \ref{Bif_point_varyD}(f)) except for small networks.

We also show the error in approximating the tipping point for Holme-Kim networks with different numbers of nodes in Figs.~\ref{Bif_point_varyD}(g) and \ref{Bif_point_varyD}(h). The DBMF theory locates the tipping point with a higher accuracy than the GBB reduction with the tipping point being defined by $\tilde{D}_{1.5}$ (Fig.~\ref{Bif_point_varyD}(g)), whereas the opposite is the case for the combination of $\tilde{D}_{2.5}$ and large $N$ (Fig.~\ref{Bif_point_varyD}(h)). All these results are similar to those when we varied $u$ instead of $D$ (see Section \ref{sub:vary-u}).

\subsection{Multistage transitions}\label{sub:multistage}

For the given $D$ value, the DBMF theory predicts a range of parameter value $u\in (u_{\rm c, \ell}, u_{\rm c,u})$ over which the different nodes tip from the lower to the upper state. Using Eq.~\eqref{eq:double_well_hmf} and Fig.~\ref{cubic_poly_ss}, we obtain
\begin{equation}
u_{\rm c, \ell} = \tilde{y}^{(1)} - D k_{\max} \Theta^{*, \ell}
\label{eq:u_min}
\end{equation}
and 
\begin{equation}
u_{\rm c, u} = \tilde{y}^{(1)} -D k_{\min} \Theta^{*, u},
\label{eq:u_max}
\end{equation}
where
\begin{equation}
\Theta^{*, \ell} = \frac{1}{N\langle k\rangle} \sum_k k p(k) x^{\rm \ell}(k)
\end{equation}
and
\begin{equation}
\Theta^{*, u} = \frac{1}{N\langle k\rangle} \sum_k k p(k) x^{\rm u}(k).
\end{equation}
We obtain $k_{\max} \gg k_{\min}$ in most networks. In contrast, the values of $\Theta^{*, \ell}$ and $\Theta^{*, u}$ are comparable although $\Theta^{*, \ell} < \Theta^{*, u}$. Therefore, except when $k_{\max}$ and $k_{\min}$ are close to each other, including the case of regular networks (i.e., those in which all nodes have the same degree), one obtains $u_{\rm c, \ell} < u_{\rm c, u}$ such that there is a nonvanishing range of $u$ in which some nodes are in the lower state and the other nodes are in the upper state. Therefore, the entire $N$-dimensional dynamical system on the network does not show a single tipping point but undergoes multiple tipping points in uncorrelated degree-heterogeneous networks even when $N\to\infty$. We refer to this phenomenon as multistage transition.
Equations~\eqref{eq:u_min} and \eqref{eq:u_max} indicate that the range of $u$ through which a multistage transition occurs
is wider for a more heterogeneous degree distribution. Similarly, when one fixes $u$ and varies $D$, a multistage transition occurs over
$D \in \left( (\tilde{y}^{(1)} - u)/(k_{\max} \Theta^{*, \ell}), (\tilde{y}^{(1)} - u)/(k_{\min} \Theta^{*, u})\right)$. 

\begin{figure}
\centering
\includegraphics[height=!,width=0.99\textwidth]{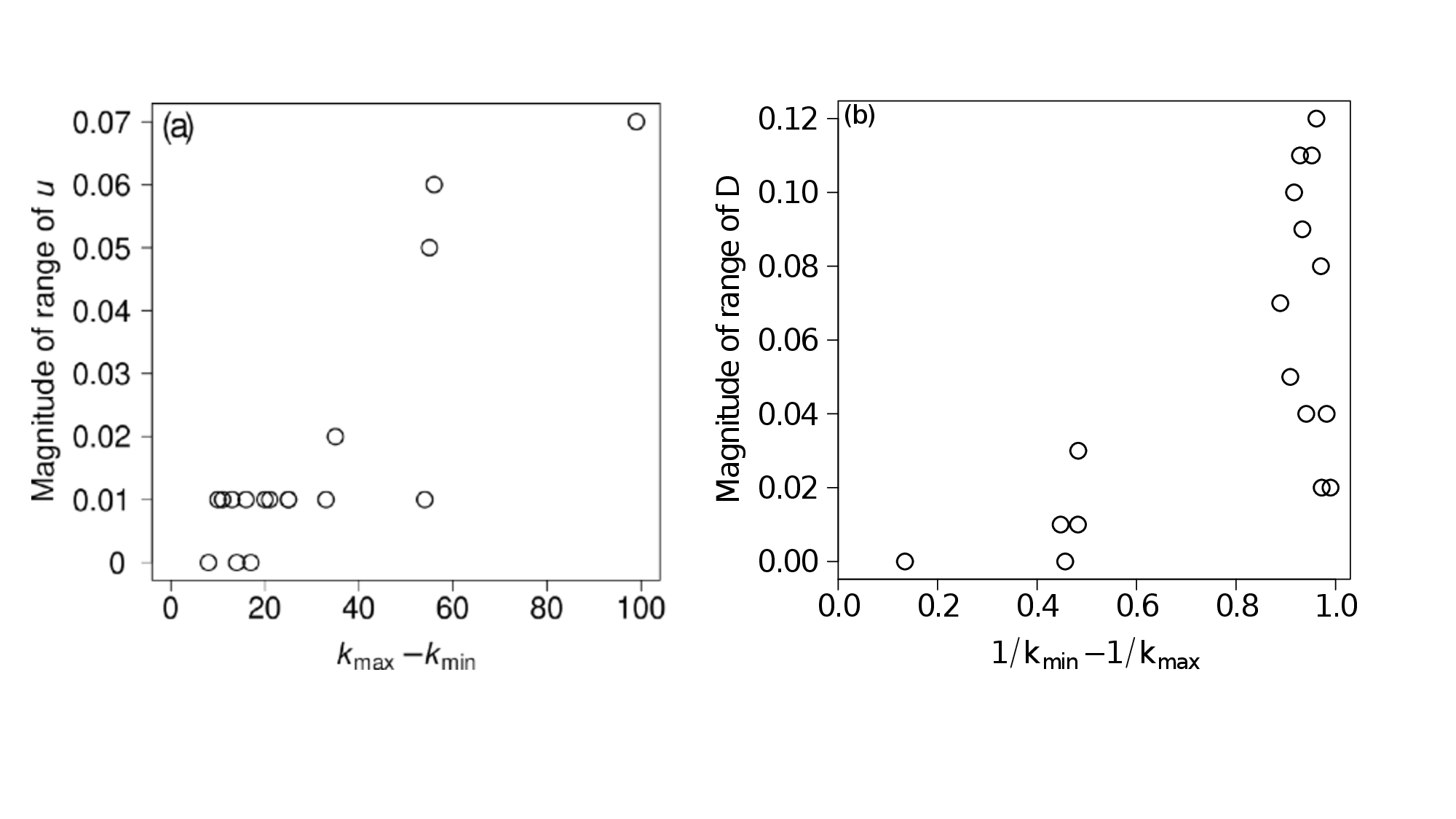}
\caption{The relationship between the degree heterogeneity and the range of a bifurcation parameter over which multiple transitions occur.
In (a), we fix $D=0.001$ and gradually increase $u$. In (b), we fix $u=0$ and gradually increase $D$. A circle represents a network. We gathered networks from \cite{schoch2022} which originated in published research \cite{silvis2014, lusseau2003, weeks2002, casey2015, freeman1998, coleman1964, adelman2015, gleiser2003, zachary1977, bull2012, firth2015, Newman_PRE2006, gill2014, freeman1988, sah2016, davis2015, vandijk2014}.
} 
\label{multistage_transition}
\end{figure}

To verify this prediction, we numerically simulated double-well dynamics on several networks. We selected 17 networks from the archive of empirical networks in the ``networkdata'' R package \cite{schoch2022} that were a mix of human and animal social networks with between 34 and 379 nodes (and mean $= 116.2$). We simplified the networks by making them undirected and unweighted, retaining only the largest connected component, and allowing no self-loops. When several networks drawn from a published work were available, we used a network with close to 100 nodes. 
%
We kept increasing the bifurcation parameter (i.e., $u$ or $D$) until at least 90\% of nodes were in the upper state at equilibrium.
Then, we calculated the range of the bifurcation parameter over which node transitions occurred as the difference between the first value when at least 10\% of nodes were in the upper state and the first value when at least 90\% of nodes were in the upper state.

In Fig.~\ref{multistage_transition}(a), we show the relationship between the observed range of $u$ and the range of node degrees (i.e., $k_{\max} - k_{\min}$) for each network when we fix $D=0.001$ and vary $u$. Each circle represents a network. Equations~\eqref{eq:u_min} and \eqref{eq:u_max} imply that the size of the range of $u$ is roughly proportional to $k_{\max} - k_{\min}$, which the results shown in Fig.~\ref{multistage_transition}(a) support. The Pearson correlation coefficient between the size of the range of $u$ and $k_{\max} - k_{\min}$ is 0.87.
In Fig. \ref{multistage_transition}(b), we show the relationship between the range of $D$ over which multiple transitions occur and $1/k_{\min} - 1/k_{\max}$ when we fix $u=0$ and vary $D$. Note that $D \in \left( (\tilde{y}^{(1)} - u)/(k_{\max} \Theta^{*, \ell}), (\tilde{y}^{(1)} - u)/(k_{\min} \Theta^{*, u}) \right)$ implies that the size of the range of $D$ is roughly proportional to $1/k_{\min} - 1/k_{\max}$. Figure~\ref{multistage_transition}(b) is also supportive of the DBMF theory; the Pearson correlation coefficient between the two quantities is 0.59. In both cases, the heterogeneity in the degree distribution in terms of $k_{\min}$ and $k_{\max}$ predicts the presence of multistage transitions fairly well.

\section{Discussion}

We have developed a DBMF theory for a standard double-well system on networks. The DBMF theory has turned out to be more accurate than the GBB reduction in locating the tipping point if we define it as the bifurcation parameter value at which a small fraction of nodes has tipped. We have also shown that the DBMF theory is less accurate than the GBB reduction at approximating the effective state substantially above the tipping point. 

The original spectral method is a generalization of the GBB reduction and uses the leading eigenvector of the adjacency matrix to linearly combine
$\{x_1, \ldots, x_N\}$ and realize a low-dimensional reduction of a family of dynamical systems on networks \cite{Laurence_PRX2019, Thibeault_PRR2020}. 
Similar to the GBB reduction, the original spectral method is accurate at approximating the effective state, whereas it is not accurate at locating the tipping point \cite{Kundu_PRE2022}. We recently extended this spectral method to the case of the nonleading eigenvector that minimizes a mathematically derived error \cite{Masuda_PRR2022}. By design, this spectral method, which we call the modified spectral method, is more accurate than the original spectral method, and presumably than the GBB reduction and DBMF theory, at approximating the effective state except near the tipping point.
However, the modified spectral method is far less accurate at locating the tipping point than the original spectral method, and also apparently than
the GBB reduction and DBMF theory \cite{Masuda_PRR2022}. Therefore, it seems that the accuracy at locating the tipping point and that at approximating the effective state sufficiently far from the tipping point are in a trade-off relationship. Specifically, among the methods discussed here, the DBMF theory is the most accurate at the former task, the modified spectral method is the most accurate at the latter task, and the GBB and the original spectral method are in between. Looking more closely into this trade-off, including searching a theory that more accurately locates the tipping point than the present DBMF theory, warrants future work.

There exists a critical value of the parameter at which the lower and upper equilibrium states have the same potential energy and thus have the same resilience \cite{vande2015,bel2012}. In this case, there exists the tipping point at which half nodes belong to the lower equilibrium and the other half to the upper equilibrium. With this definition of the tipping point, i.e., $u=\tilde{u}_{2.5}$ or $D=\tilde{D}_{2.5}$ in our case, we have shown that our DBMF theory is not accurate. In the present study, our primary definition of the tipping point is the value of the parameter at which a small fraction of nodes (5--10\%) has transitioned from the lower to upper state (i.e., $u=\tilde{u}_{1.5}$ or $D=\tilde{D}_{1.5}$). We chose such a small fraction because, in practice, we are often interested in the situation in which a small but noticeable fraction of nodes, rather than half the nodes, has tipped to enter an undesirable state. For example, the tipping point for the flip to non-forest ecosystems in eastern, southern and central Amazonia is estimated to be at 20--25\% deforestation \cite{lovejoy2018}. We have shown that our DBMF theory is more accurate than the GBB reduction with this latter definition of tipping.

 In the present study, we have assumed that the influence of the $j$th node on the $i$th node is proportional to $A_{ij}x_j$, where $A_{ij}$ is the $(i, j)$ element of the adjacency matrix. In contrast, double-well systems on networks in which the nodes are diffusively coupled have also been
studied. Such a model represents, for example, chemical reaction diffusion
\cite{Kouvaris_PlosOne2012, Kouvaris_EPL2013,Yang_EPJB2014}. A DBMF theory was previously developed for the case of diffusive coupling~\cite{Kouvaris_PlosOne2012}. However, adapting their theory to be able to self-consistently determine the mean field (i.e., $\Theta^*$ in our formulation) and the threshold degree (i.e., $\tilde{k}$), and thus further pursuing how nodes with different degrees behave and comparing the results with the case of coupling by the adjacency matrix warrants future work.

We have provided theoretical and numerical evidence of multistage transitions in the present double-well system on networks. In particular, our DBMF theory suggests that the multistage transition is not a finite-size effect because the DBMF theory is more accurate in large uncorrelated networks in general. In fact, our DBMF theory
depends on the number of nodes, $N$, through Eq.~\eqref{eq:self_consistenta_Theta^*}. However, an increase in $N$ does not mitigate the validity of Eqs.~\eqref{eq:u_min} and \eqref{eq:u_max}, which suggests that multistage transitions are likely to occur in networks of different sizes. Investigating multistage transitions for a wider variety of networks and dynamical systems, as well as their practical implications, also warrants future work.

\section*{Acknowledgments}

H.K. acknowledges support from JSPS KAKENHI under Grant No. 21K12056. N.M. acknowledges support from AFOSR European Office (under Grant No. FA9550-19-1- 7024), the Sumitomo Foundation, the Japan Science and Technology Agency (JST) Moonshot R\&D (under Grant No. JPMJMS2021), and the National Science Foundation (under Grant No. 2052720).



\end{document}